\documentclass[useAMS,usenatbib]{mn2e}

\usepackage{amsmath}
\usepackage{comment}
\usepackage{graphicx}
\usepackage{rotating}
\usepackage{color}
\usepackage{aas_macros}
\usepackage{amsfonts}
\newcommand{\beq}{\begin{equation}}
\newcommand{\enq}{\end{equation}}
\newcommand{\beqa}{\begin{eqnarray}}
\newcommand{\enqa}{\end{eqnarray}}
\newcommand{\beit}{\begin{itemize}}
\newcommand{\enit}{\end{itemize}}
\newcommand{\bem}{\begin{pmatrix}}
\newcommand{\enm}{\end{pmatrix}}
\newcommand{\veck}{\mathbf{k} }

\newcommand{\lp}{\left (}
\newcommand{\rp}{\right )}
\newcommand{\bes}{\begin{sideways}}
\newcommand{\ees}{\end{sideways}}

\renewcommand{\bbeta}{\boldsymbol {\eta}}
\renewcommand{\bmu}{\boldsymbol {\mu}}

  \title[On the total information]{On the total cosmological
  information in galaxy clustering: an analytical approach}

\author[Wolk, Carron and Szapudi]{M. Wolk\thanks{E-mail:
wolk@ifa.hawaii.edu}, J. Carron and I. Szapudi  \\
Institute for Astronomy, University of Hawaii, 2680 Woodlawn Drive, Honolulu, HI, 96822}
\begin{document}

\date{\today}

\pagerange{\pageref{firstpage}--\pageref{lastpage}} \pubyear{2014}

\maketitle

\label{firstpage}

\begin{abstract}
%Beyond the linear regime, the cosmological information contained in
%the galaxy clustering is erased from both the power spectrum and the
%higher order moments due to correlation between Fourier modes induced
%by non-linear gravitational evolution. The ``sufficient statistics'' $A^*$ was introduced recently
%as a simple mean to recapture as efficiently as possible this lost
%information into its spectrum. Even though, already been proven successful,
%this new data analysis strategy involved numerical simulations of both
%the power spectra and covariance matrices hiding the physics of the
%non-linear transform. In this work, we propose an
%analytical approach to describe the $A^{\ast}$-power spectrum and its
%covariance matrix thus allowing insights about the total
%information content of the galaxy clustering. In order to compare to the efficiency of the
%widely used observable, the galaxy power spectrum, we also introduce
%an analytical approximation of its covariance in the case of projected
%fields. Using these ingredients to quantify the 
%cosmological information content of the galaxy field, we derive the
%best achievable constraints one could expect as a function of the survey
%shot-noise level. We also provide an estimation of the
%gain in our knowledge of cosmological
%parameters obtained using the new observable $A^{\ast}$ instead of the
%galaxy power spectrum and find a factor $\sim 2$ for dense
%low-redshift surveys, recovering the previous results from numerical simulations.
Beyond the linear regime of structure formation, part of cosmological information encoded in galaxy clustering becomes inaccessible to the usual power spectrum.
{\em Sufficient statistics}, $A^*$, were introduced recently to recapture the lost, and ultimately extract all, cosmological  information.
We present analytical  approximations for the $A^{\ast}$ and
traditional power spectra as well as for their covariance matrices in
order to calculate analytically their cosmological information content
in the context of Fisher information theory. Our approach allows the
precise quantitative comparison of the techniques with each other and
to the total information in the data, and provides insights into
sufficient statistics. In particular, we find that while the $A^*$
power spectrum has a similar shape to the usual galaxy power spectrum, its amplitude is strongly modulated by small scale statistics. This effect is mostly responsible for the ability of the  $A^*$ power spectrum to recapture the information lost for the usual power spectrum.
We use our framework to forecast the best achievable cosmological constraints for projected surveys as a function of their galaxy density, and compare the information content of the two power spectra. We find that sufficient statistics extract all cosmological information, resulting in an approximately factor of $\simeq 2$ gain for dense projected surveys at low redshift.
This increase in the effective volume of projected surveys is consistent with previous numerical calculations.

\end{abstract}


\section{Introduction}
Within the current inflationary paradigm of cosmology, the small
initial density fluctuations are believed to be very close to Gaussian
statistics. The most natural observables of such a field, the two-point
statistics, lose some of their statistical power as non-linear gravitational growth induces
correlations between Fourier modes \citep{Rimesetal05, Neyrincketal06}. 
These correlations, especially those between large and small scales, diminish the amount of information accessible to these two-point statistics. A fraction of this hidden
information is accessible to higher-order $N$-point statistics
\citep[e.g.,][]{Peebles1980, Szapudi2009}. They are, however, not only 
difficult to measure and interpret due to a combinatorial explosion of complexity, but they fail to capture all available cosmological information, increasingly so on more non-linear scales. \citep[and references therein]{CarronNeyrinck2012,Carronetal13}.

Non-linear transformations, such as the logarithmic
mapping \citep{Neyrincketal09} or variants thereof
\citep{Seoetal11,Joachimietal11} were introduced specifically to retrieve the total
information content of the matter field. \cite{Carronetal13} defined {\em sufficient statistics} as an observable extracting all cosmological information from data. They
have demonstrated in the context
of perturbation theory and $N$-body simulations that the logarithmic
transformation, $A = \ln(1 + \delta)$, approximates well the exact
sufficient statistics of the dark matter field. Note that in the case of a continuous lognormal
field $A$ is the exact sufficient statistics, a statement supported by analytical calculations and measurements in simulations \citep[][and references therein]{Carronetal14c}.
\newline
In a previous work, \cite{Carronetal14a}  introduced the local 
non-linear transformation $A^*$ as the optimal observable to extract the
information content of galaxy count maps.
This recaptures in its spectrum the total available cosmological 
information in presence of shot-noise. The new observable has been characterized in
detail using numerical simulations  of
2-dimensional survey configurations \citep{Wolketal14, Carronetal14c}. 
Yet, the precise manner in which $A^{\ast}$
recaptures the cosmological information remained somewhat of a puzzle, given that it's shape closely resembles that of the power spectrum.
In this work, we present an analytical theory of the total information content (i.e. the constraining
power) of the $A^*$ angular power spectrum for cosmological
parameters, assuming that we have access to the power spectrum and its
derivatives as a function of cosmological parameters; this is provided
by a standard Boltzman code, such as CAMB \citep{Lewis1999bs}. Our approach is then used to 
compare sufficient statistics with the usual angular galaxy power spectrum as a
function of the relevant projected survey characteristics, most importantly the shot noise level. The analytical approach provides insight into the workings of sufficient statistics, in particular
it sheds light on the crucial role played by the bias of the non-linear transformation in recapturing the lost information.
\newline
To test its validity, we carefully compare our model to the predictions
from our previous numerical simulations. For simplicity of expression, we will designate these numerical results as  ``exact'' throughout this work. In practice, these simulations provide accurate enough results that this nomenclature is justified.
Throughout this paper, the notation $P(k)$ is used to designate the angular
power spectrum with $k \simeq \ell + 1/2$ in the flat sky approximation.

Section~\ref{sec:Astar} describes the
analytical ansatz for the $A^{\ast}$ bias and covariance matrix. Described in Section~\ref{sec:covfield} is the
model for the 2-dimensional matter field covariance matrix.  
With this model, we estimate the information content for different
survey densities as presented in Section~\ref{sec:infocontent}. Our
estimations are also compared to previous numerical
predictions of \cite{Wolketal14} . We summarize and conclude with a discussion in Section~\ref{sec:discuss}. 

\section{Ansatz for the $A^{\ast}$ bias and covariance matrix}
\label{sec:Astar}
Assuming that the galaxy counts  Poisson sample an underlying
lognormal galaxy field, let $N =
(N_1,\cdots,N_{n_{cells}})$ be a map of galaxy counts. In the
following, $n_{cells}=128^{2}$, for a two dimensional map. 
Given a sampling rate $\bar N$, the mapping from $N$ to $A^*$ is defined by the non-linear equation \citep{Carronetal14a}:
\beq \label{Astar}
A^* + \bar N \sigma^2_{\ast} e^{A^*}   = \sigma^2_{\ast} \lp N - \frac 12\rp,
\enq
where $\sigma^2_{\ast} = \ln (1 + \sigma^2_{\delta_g})$, with
$\sigma^2_{\delta_g}$ the variance of the galaxy field fluctuations at
the cell scale. These $A^{\ast}$-mapping parameters are estimated
using our fiducial cosmology and are then kept fixed. Given the
current precision of cosmological parameters this assumption amounts 
to no practical limitations for our technique.

\cite{Wolketal14}, using a simulation pipeline calibrated on the
Canada-France-Hawaii Telescope Legacy Survey (CFHTLS)\footnote{\texttt{http://www.cfht.hawaii.edu/Science/CFHTLS/T0007/}}
data, have shown that the information
gain using the mean and spectrum of $A^{\ast}$ instead of the galaxy
power spectrum on the three cosmological parameters $\Omega_{m}$,
$\sigma_{8}$ and $w_{0}$ is up to about a factor of 2, especially at low
redshifts and for dense surveys.
This numerical approach clearly demonstrated that the
``sufficient statistics'' $A^{\ast}$ performs better, yet, it could not yield qualitative
insights into workings of sufficient statistics. Given that the shape of the $A^{\ast}$
power spectrum is very similar to the usual power spectrum the question naturally arises:
is the increase of information attributed to $A^{\ast}$ itself, more
precisely its derivatives being more sensitive to parameters, or, to
the fact that the corresponding covariance matrix is better behaved,
in particular more diagonal? In our analytical approach next we point
out the crucial cosmological dependence of the bias, and show that part of the information gain in fact can be pinned on the derivatives of the bias with respect to cosmological parameters.

\subsection{From the galaxy power spectrum to the $A$-power spectrum}
\begin{figure}
  \begin{center}
    \includegraphics[width=9cm]{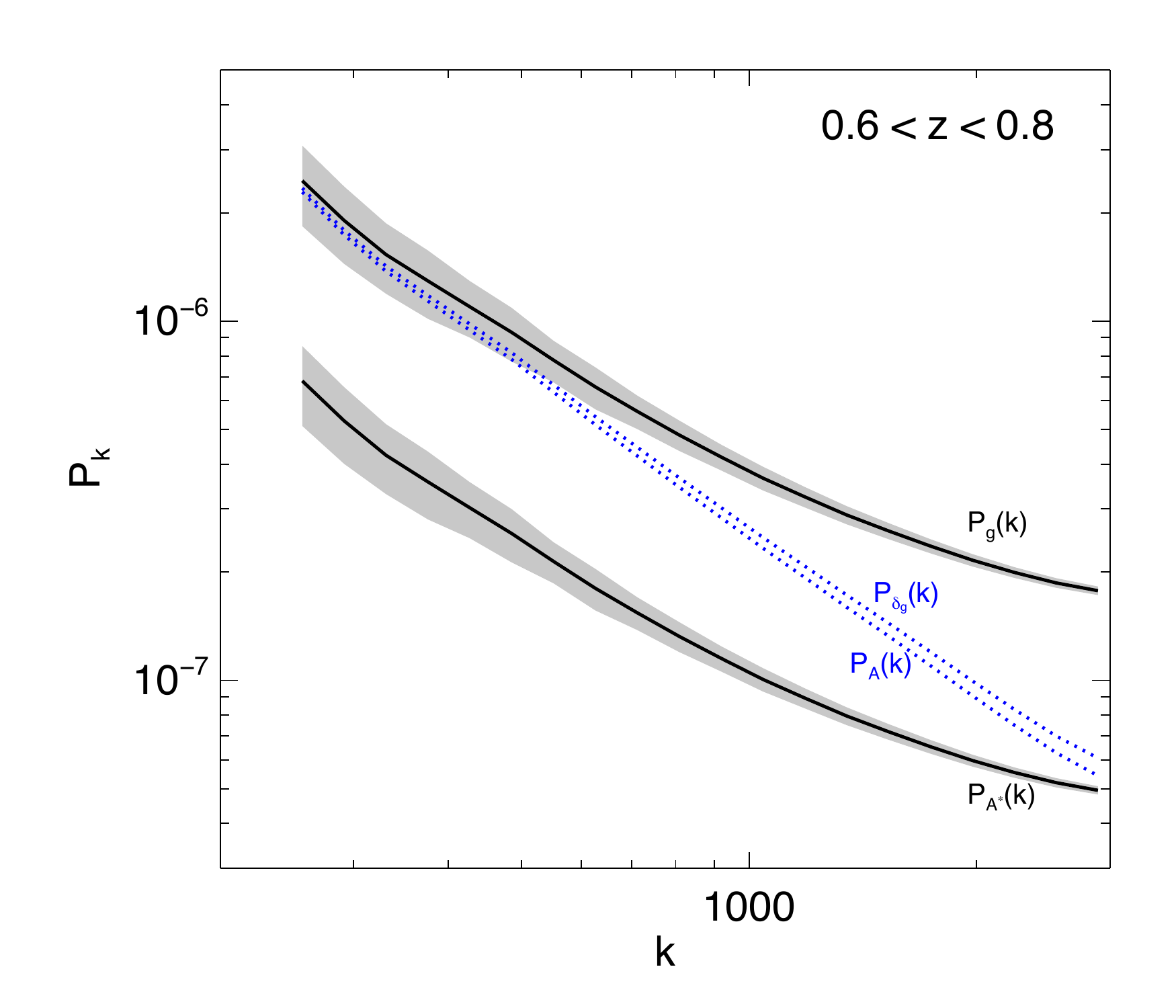}
\end{center}
\caption{Predictions of both the galaxy angular power spectrum
  $P_{g}(k)$ and the non-linear transform $A^{\ast}$ power spectrum $P_{A^{\ast}}(k)$
  in the redshift bin $0.6<z<0.8$. The grey area show the 1$\sigma$
  confidence regions. The blue dotted lines represent the power spectra of
  the underlying fields $\delta_{g}$ and $A=\ln(1+\delta)$ and thus
  illustrate the effect of shot-noise.}
\label{fig:Pkexample} 
\end{figure}

Our analytical approach assumes prior knowledge on the galaxy power
spectrum $P_{\delta_{g}} \equiv P$ as function of cosmological (and halo) parameters.
We use the standard, unweighted power spectrum estimator:
\beq
\hat{P}(k) = \frac{1}{VN_{k}}\sum_{k'}|\delta(k')|^{2}
\enq
with $V$ the survey volume and where the sum runs over the $N_{k}$
Fourier modes associated to the $k$-th power spectrum bin. This simple estimator is optimal for simple geometries, such as $N$-body simulations. Including complications from survey geometry and the corresponding optimal weighting of the estimator will not change any of our results, as the scales we are focussing on are small enough that edge effects will become unimportant.
We model the galaxy clustering with the Halo Occupation Distribution (HOD) description of \cite{Wolketal14} and the
\texttt{CosmoPMC}\footnote{\texttt{http://cosmopmc.info}} package. We
consider four different redshift bins: $0.2<z<0.4$, $0.4<z<0.6$,
$0.6<z<0.8$ and $0.8<z<1.0$.
Figure~\ref{fig:Pkexample} shows the $A^{\ast}$- and galaxy power spectra for the
redshift bin $0.6<z<0.8$ with their 1$\sigma$
confidence regions (shaded) as well as the predictions of the power spectra for the
underlying fields $\delta_{g}$ and $A=\ln(1+\delta_{g})$ (blue dotted
lines).

The first step is to model the bias between the spectra of the two continuous
fields $\delta_{g}$ and $A = \ln(1 + \delta_{g})$.
Here we assume a lognormal underlying
galaxy density, an hypothesis that have been proved to be very
accurate in 2D \citep{Carronetal14c}.
Then the simplest Ansatz to consider is the ratio of the
variances, and for the lognormal model the variances are related as
$\sigma_{A}^{2} = \ln(1 + \sigma_{\delta_{g}}^{2})$. Explicitly, we assume
\beq
P_{A} = b_{A}^{2} P
\label{eq:biascont}
\enq
where:
\beq
b_{A}^{2} = \frac{\sigma_{A}^{2}}{e^{\sigma_{A}^{2}} -1}.
\enq

According to the left panel of Figure~\ref{figure:biascont}, this approximation is better 
than $4\%$ accurate on all the $k$-range, even
if it starts to deviate slightly both for very large or very small scales. This formula, in the regime of low
$A$-variance, reduces to that of \cite{Neyrincketal09}
$b_{A}^{2}=e^{-\sigma_{A}^{2}}$ obtained from 
simulations for the 3-dimensional power spectrum.

\subsection{From continuous to discrete fields}

The link between the continuous and discrete galaxy field power
spectra is well understood for Poisson sampling through $P_{g} = P + 1/\bar{n}$
where $\bar{n}$ is the density of the considered survey related to the
sampling rate via $\bar{n} = \bar{N} n_{cells}/V$. $V$ is the
survey volume and is fixed here to the size of the CFHTLS-W1 field $L = 7.46$
degrees on the side. As it will be explained in more details in
Section~\ref{sec:covfield}, considering the local lognormal case will
result in a cancellation of the contribution from the super survey
modes in the galaxy power spectrum covariance matrix, hence implying
that the information content does not depend on
the survey geometry.
Then, how could the relationship be explained between the $A$- and the
$A^{\ast}$ power spectra? 

\begin{figure*}
  \begin{center}
  \begin{tabular}{c@{}c@{}}
      \includegraphics[width=0.45\textwidth]{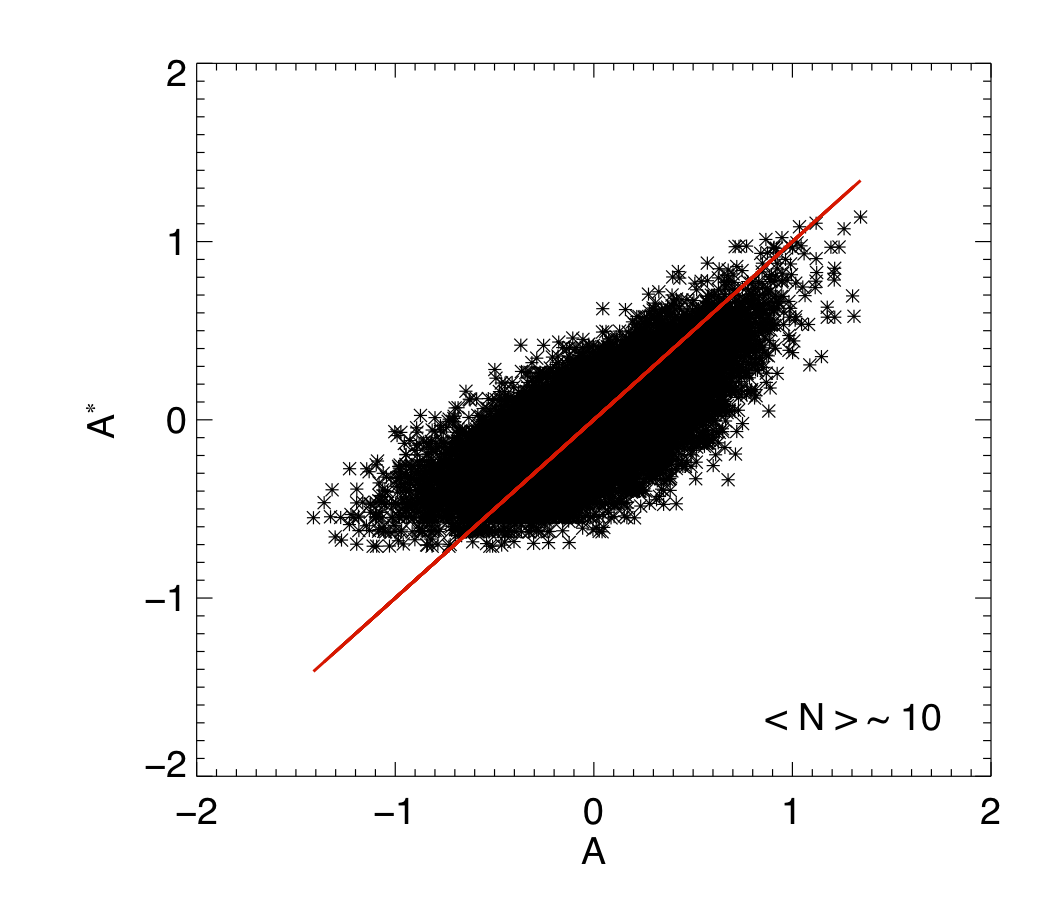}&
      \includegraphics[width=0.45\textwidth]{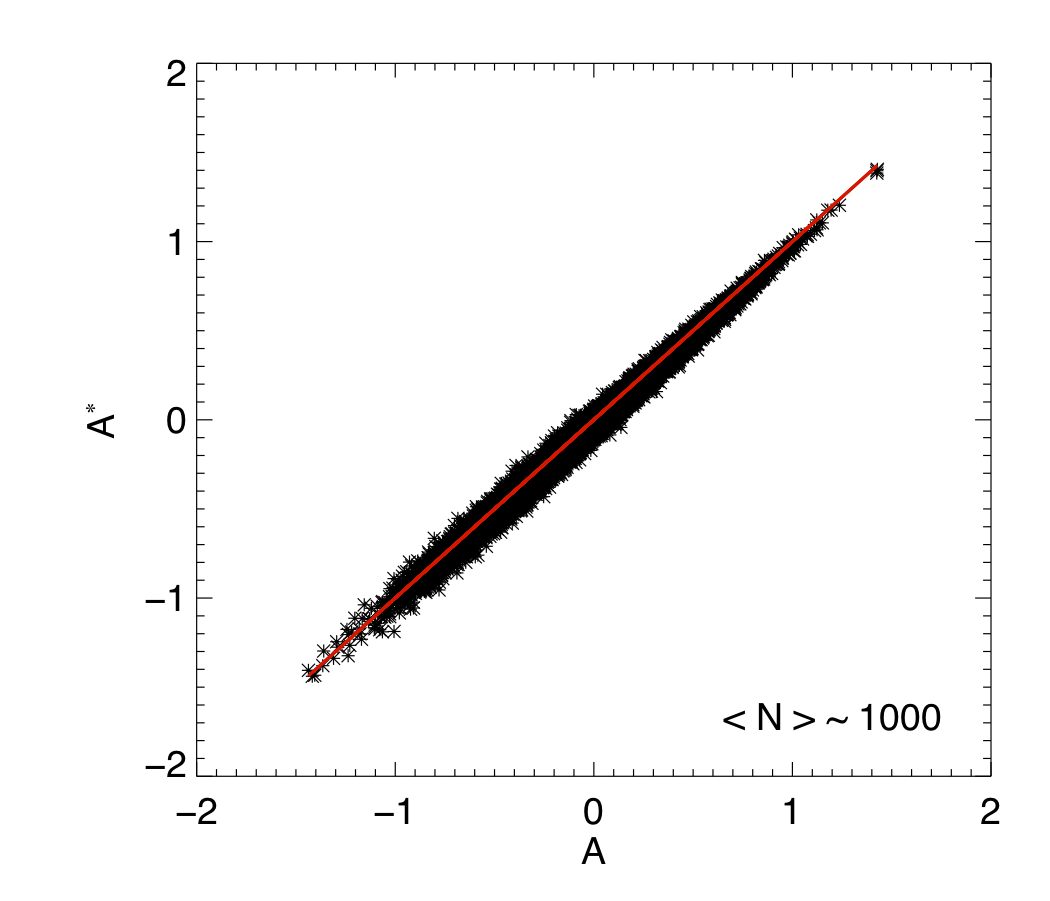}
\end{tabular}
   
    \caption{ \label{fig:scatter} Scatter plot of $A^{\ast}$ as a function of A for two
      different values of $\bar{N}$ using $n_{cells}=128^{2}$. The red
    lines represents the cell values for which $A^{\ast}=A$. When the survey is dense, i.e when $\bar{N}$ is
large (right panel), there is almost no bias between $A$ and
$A^{\ast}$. However, for low density survey (left panel), $A^{\ast}$ tends to be smaller than
$A$ leading to a bias between the two quantities.}
\end{center}
\end{figure*}

From Equation~\ref{Astar}, it can expected that
this relationship depends on the $A^{\ast}$-mapping parameters and
especially on $\bar{N}$ as, at
a particular redshift and $n_{cells}$, $\sigma_{\ast}$ is fixed.
Figure~\ref{fig:scatter} shows the scatter plot of $A^{\ast}$ as a
function of $A$ for two different values of $\bar{N}$, the first
one corresponding to the sampling rate of the CFHTLS-W1 field in the
redshift bin $0.6<z<0.8$.

When the survey is dense, i.e when $\bar{N}$ is
large enough, there is almost no bias between $A$ and $A^{\ast}$
meaning that the local transformation $A^{\ast}$ traces well the
underlying field. In contrast, for a low density survey, $A^{\ast}$ tends to be smaller than
$A$ leading to less fluctuations thus less power in $A^{\ast}$ in
agreement with the simulations on Figure~\ref{fig:Pkexample}. Hence most of the bias is due to the fact 
that, for low $\bar{N}$, $A^{\ast}$ cannot distinguish between low
$A$ regions or a cell that happens to be empty due to a low $\bar{N}$
\citep[a non-local generalization of $A^{\ast}$ would potentially behave better][]{Carronetal13}.

\begin{figure*}
 \begin{center}
  \includegraphics[width=0.45\textwidth]{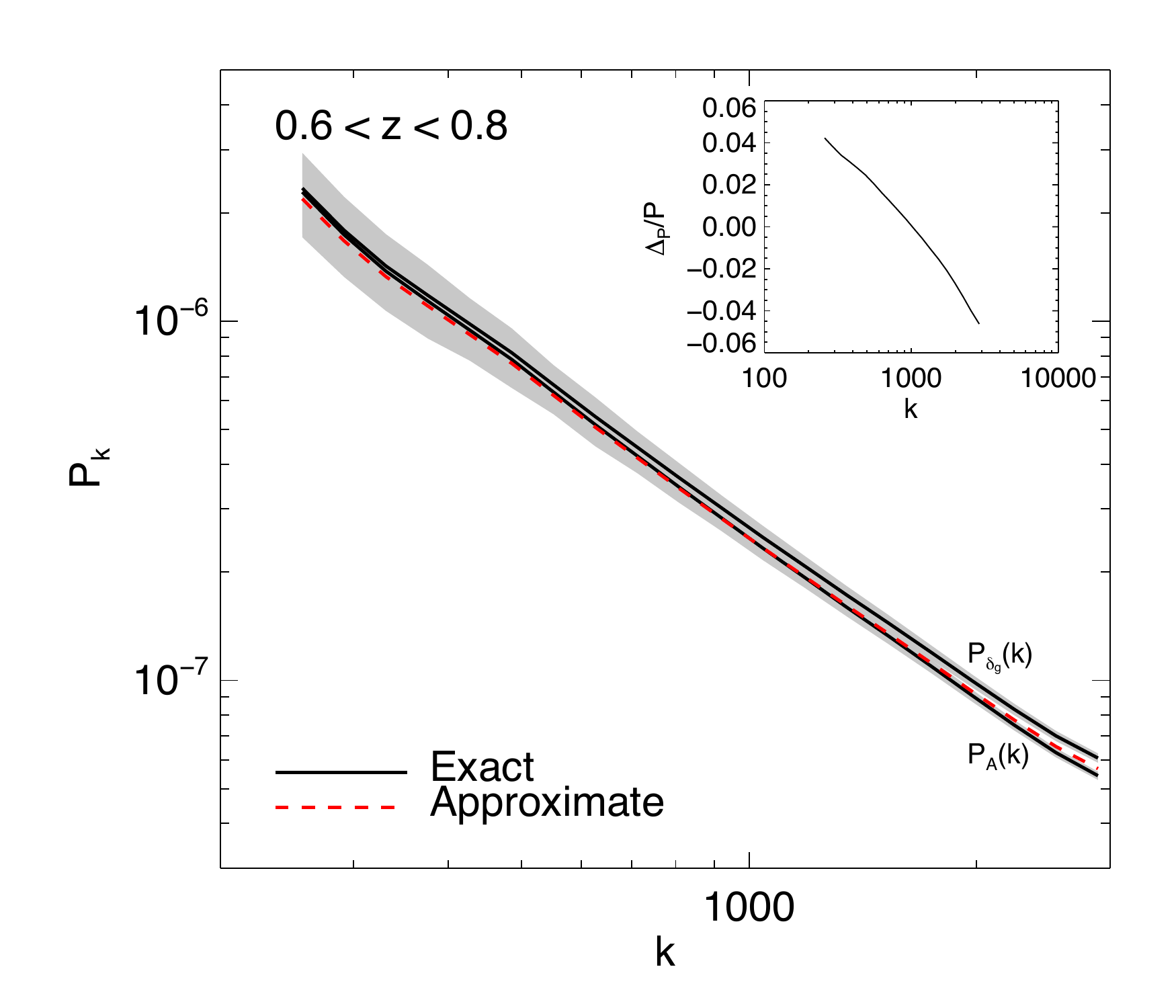}
  \includegraphics[width=0.45\textwidth]{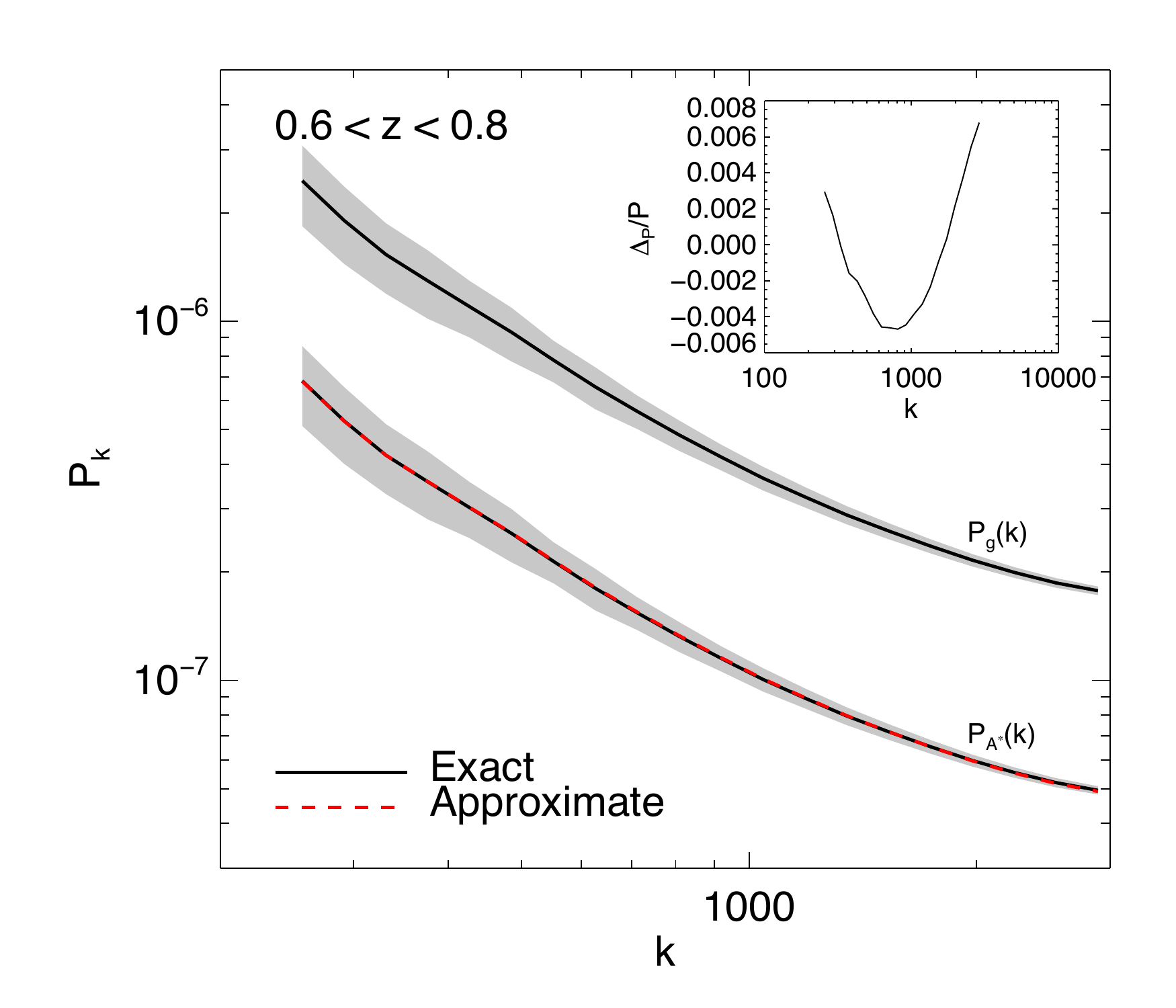}
   \caption{ \label{figure:biascont} On the left panel, the solid
     black lines represent the prediction of power spectra of
  the underlying fields $\delta$ and $A=\ln(1+\delta)$ in the
      redshift bin $0.6<z<0.8$. The dashed red line is the result
      obtained using Equation~\ref{eq:biascont}. On the right panel, predictions of both the galaxy angular power spectrum
  $P_{g}(k)$ and the non-linear transform $A^{\ast}$ power spectrum $P_{A^{\ast}}(k)$
  in the redshift bin $0.6<z<0.8$. The dashed red line is the result
      obtained using Equation~\ref{eq:biasfinal}. On both panels, the grey area show the 1$\sigma$
  confidence regions and the inside panels show the deviation from the
true value using our model.}
\end{center}
   \end{figure*}

The next step is to relate the galaxy power spectrum to the $A^{\ast}$
power spectrum. In order to take the shot-noise contribution into
account, we develop to the 2$^{\rm{nd}}$ order expansion around zero for
the exponential term of Equation~\ref{Astar} and then take the Fourier
transform:
\beq
\Biggr[\frac{1}{\sigma^2_{\ast} } + \bar N e^{\bar{A^*}} \Biggl]^{2} P_{A^*} = P_{N} =
\bar{N}^{2} P_{g} = \bar{N}^{2} \Big[ P + \frac{1}{\bar{n}} \Big].
\enq
Thus the bias has a simple form:
\beq
P_{A^{\ast}} =  \frac{b_{A}^{2}}{b_{A^{\ast}}^{2}} P_{g} = \frac{b_{A}^{2}}{b_{A^{\ast}}^{2}}\Big[ P + \frac{1}{\bar{n}} \Big]
\label{eq:biasfinal}
\enq
where
\beq
b_{A^{\ast}}^{2} = \Big(1+\frac{1}{\bar{N}\sigma_{\ast}^{2}} \Big)^{2}
\enq
The right panel of Figure~\ref{figure:biascont} shows the $A^{\ast}$-
and galaxy power spectra as well as the prediction from
Equation~\ref{eq:biasfinal}. The accuracy is better than $0.5\%$ over
the whole $k$-range.

\subsection{The $A^{\ast}$ covariance matrix}
To quantify its Fisher information content, we need an estimation of
the $A^{\ast}$-covariance matrix:
\beq
\textrm{Cov}^{A^{\ast}}_{ij} = \langle \hat{P}_{A^{\ast}}(k_i)
\hat{P}_{A^{\ast}}(k_j) \rangle - \langle \hat{P}_{A^{\ast}}(k_i) \rangle \langle \hat{P}_{A^{\ast}}(k_j) \rangle
\enq
We found previously that a diagonal Gaussian covariance provides an accurate model:
\beq
\textrm{Cov}^{A^{\ast}}_{ij} = \frac{2}{N_{k}} P_{A^{\ast}}(k_i)
P_{A^{\ast}}(k_j)\delta_{ij}, 
\label{eq:covAs}
\enq
where the $A^{\ast}$-power spectrum is given by Equation~\ref{eq:biasfinal}.
This is further motivated by the fact that i) \cite{Carronetal13} have shown
non-linear transformations tend to Gaussianize the field, ii) in our model of
lognormal underlying distribution, it would be exact in the absence of shot-noise (i.e when
$\bar{N}$ goes to infinity) and iii)
taking shot-noise into account tends to increase the diagonal part of the covariance
matrix adding an extra term $(1/\bar{n})^{2}$.

\begin{figure*}
 \begin{center}
 \begin{tabular}{c@{}c@{}}
      \includegraphics[width=5.5cm]{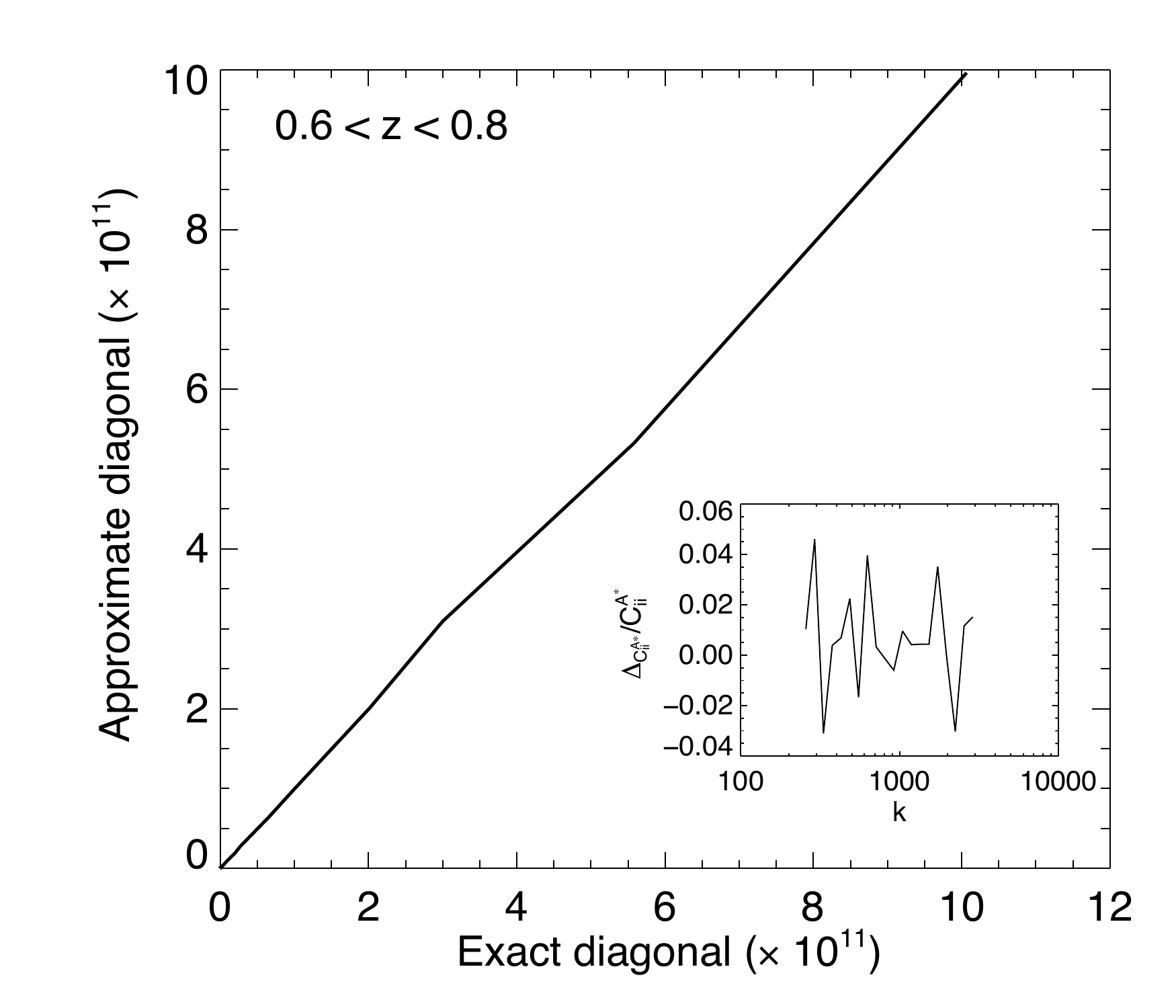}
      \includegraphics[width=5.8cm]{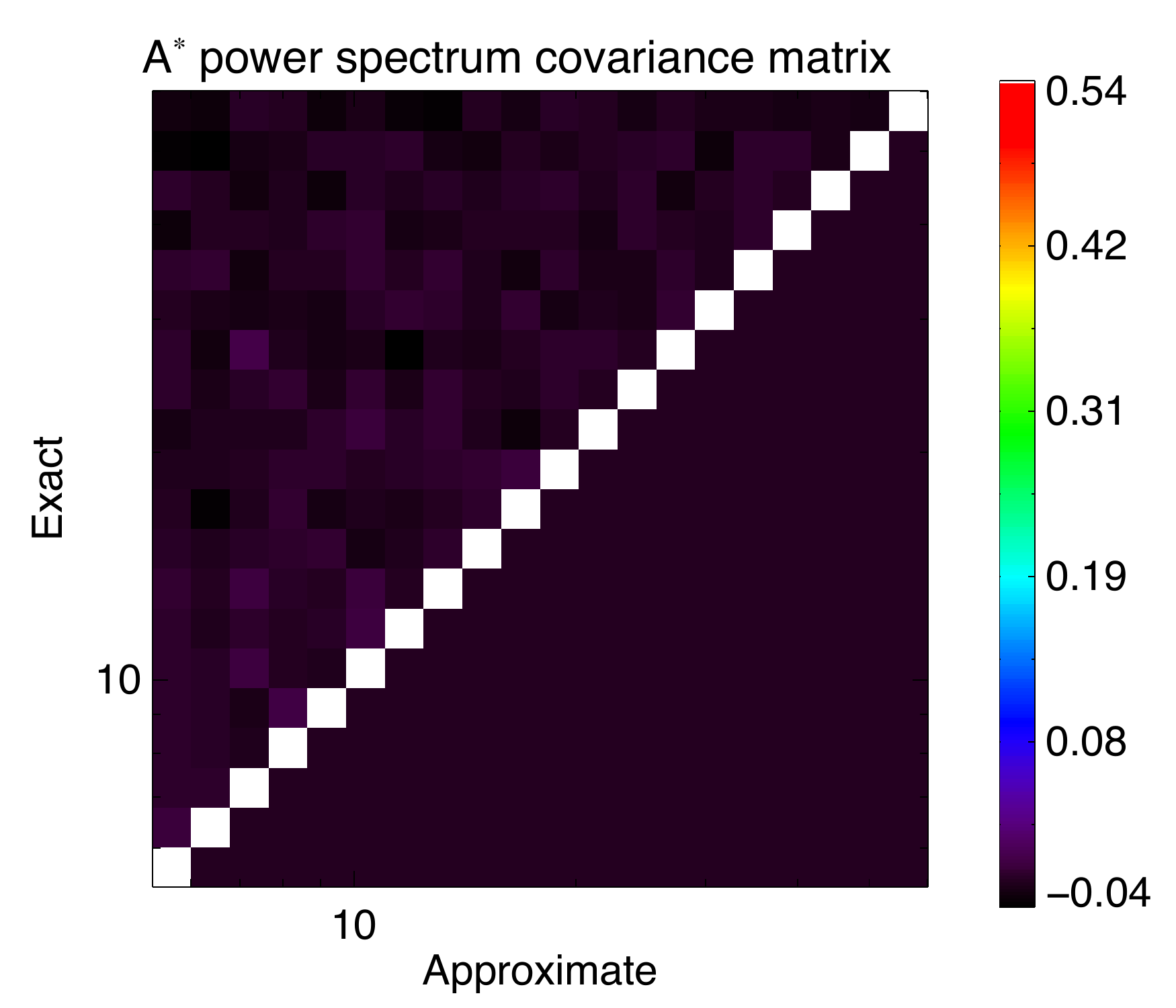}
      \includegraphics[width=5.5cm]{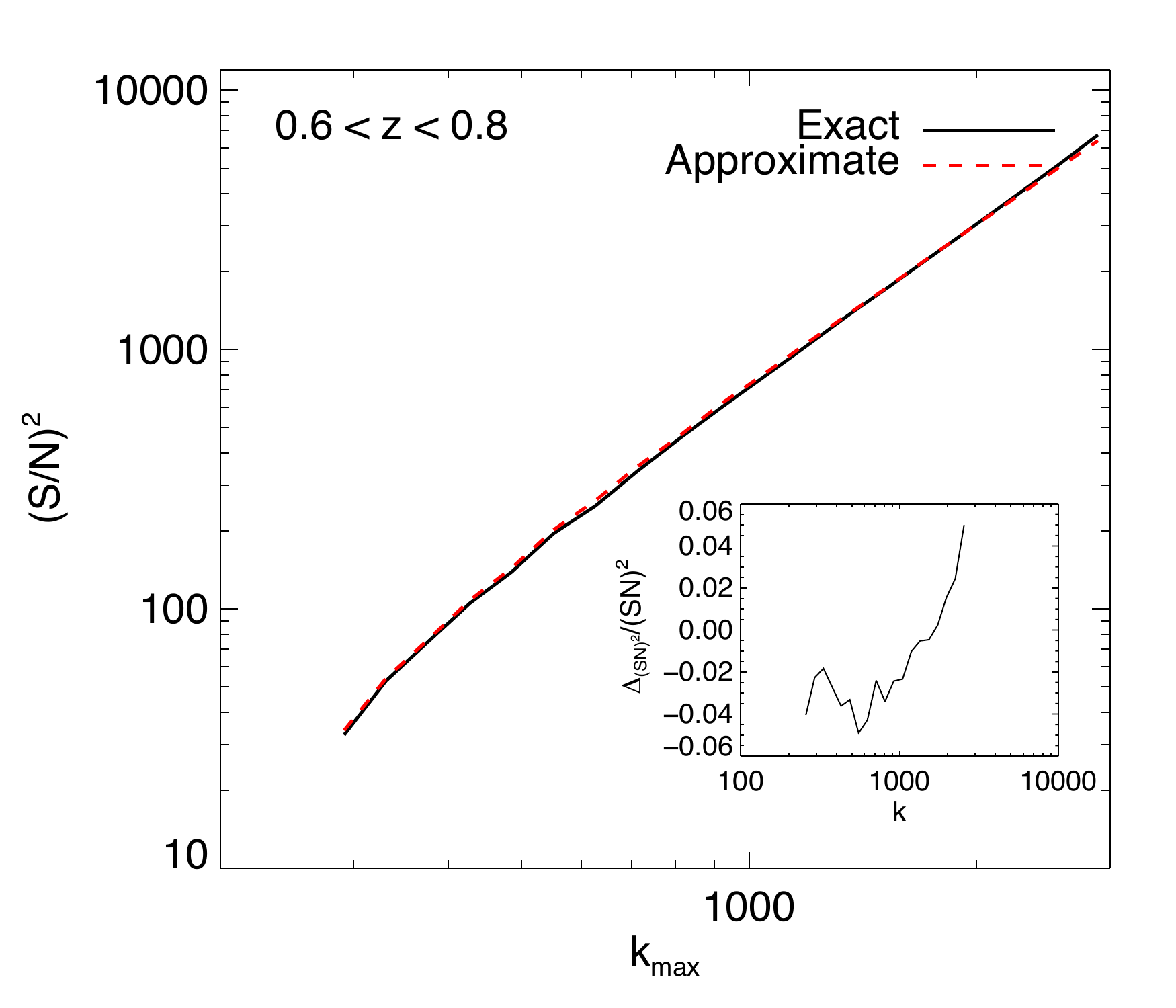}
      \end{tabular}
   \caption{ \label{figure:Ascov} On the left panel, the diagonal of
     the $A^{\ast}$-power spectrum covariance matrix obtained using Equation~\ref{eq:covAs} as a function of the
exact value obtained with simulations. The middle panel represents the comparison between
the approximate (lower right) and the exact (upper left) values of the normalised
$A^{\ast}$-covariance matrix. At $<10\%$ level, the analytical formula
reproduces very well the exact prediction. The right panel shows the squared cumulative
signal-to-noise for $A^{\ast}$ obtained using approximation of
Equation~\ref{eq:covAs} compared to the exact value. At our
resolution, the accuracy is better than $5\%$.}
\end{center}
   \end{figure*}

The left panel of Figure~\ref{figure:Ascov} shows the diagonal of the
matrix obtained using Equation~\ref{eq:covAs} as a function of the
exact value. The agreement is almost perfect on the diagonal between
the two quantities. The middle panel represents the comparison between
the approximate (lower right) and the exact (upper left) values of the normalised
$A^{\ast}$-covariance matrix. The analytical formula
reproduces the exact prediction at the $10\%$ level or better. To quantify the impact of
these discrepancies on the non-diagonal terms, we can consider the squared cumulative
signal-to-noise for $A^{\ast}$ defined as:
\beq
(S/N)^{2} = \sum_{k_i, k_j \leq k_{max}} P_{A^{\ast}}(k_{i})
  [\textrm{Cov}^{A^{\ast}}_{ij}]^{-1} P_{A^{\ast}}(k_{j})
\enq
as a function of the resolution $k_{max}$.
This is shown on the right panel of Figure~\ref{figure:Ascov}, at our
resolution $k_{max} \sim 3000$, the accuracy of
Equation~\ref{eq:covAs} is better than $5\%$.

\section{The 2D galaxy field covariance matrix}
\label{sec:covfield}
As the galaxy power spectrum is among the most widely used statistic
to extract information about cosmological parameters in large scale
structures surveys, it is worth quantifying how much
information one can expect on a given parameter as a function of the
survey characteristics and compare it to the total information on this
parameter available from the data set. 
\newline
\newline
\cite{Carronetal14d} developed a useful, approximate form of the matter power
spectrum in the mildly non-linear regime based on previous 
studies from $N$-body simulations \citep{Neyrinck11, Mohammedetal14}:
\beq \label{Cov3d}
\begin{split}
\textrm{Cov}_{ij} &= \langle \hat{P}(k_i)\hat{P}(k_j) \rangle - \langle \hat{P}(k_i)
\rangle \langle \hat{P}(k_j) \rangle \\
&= \delta_{ij} \frac{2 \lp P(k_i)+\frac 1 {\bar n}\rp^2}{N_{k_i}} +
\sigma_{min}^{2}P(k_i)P(k_j).
\end{split}
\enq
The first term corresponds to the Gaussian covariance and the second
term approximates the shell-averaged trispectrum of the field.
It turns out that the parameter $\sigma_{min}^{2}$, can be interpreted as the
minimum variance achievable on an amplitude-like parameter \citep[see][for details]{Carronetal14d}. It can be further decomposed into two contributions:
\beq
\sigma_{min}^{2} = \sigma_{SS}^{2} + \sigma_{IS}^{2}.
\enq
The first term is due to the correlation between large
wavelength ``super-survey'' modes with the small scales while the
second term corresponds to the coupling between small scales or
``intra-survey'' modes.

Here we study local density fluctuations,
$\delta=\frac{\rho-\bar{\rho}}{\rho}$, defined with respect to the
local observed density. In the particular case of a lognormal
underlying distribution, there is a cancellation between two
contributions in the covariance matrix resulting in $\sigma_{SS}^{2} =
0$ \citep[see][for details]{Carronetal14d}.
Thus in our study, the only significant contribution comes from the
``intra-survey'' modes. We model $\sigma_{min}^{2}$ within the hierarchical
\textit{Ansatz} \citep{Peebles1980, Fry84, Bernardeau96}, reducing to
\citep[see details in][]{Carronetal14d}:
\beq
\begin{split}
& \sigma_{min}^{2} = \sigma_{IS}^{2} = (4R_{a}+4R_{b})\frac{1}{\sigma_{\delta_{g}}^{2}}
\frac{1}{V} \int \frac{dk}{2\pi} kP^{2}(k) \\
& \simeq \frac{P(k_{max})}{V}(4R_{a}+4R_{b})
\end{split}
\label{eq:sigcst}
\enq
which decreases as the resolution increases.

Although it has been proved to be a good model in the 3-dimensional case, this
approximation does not
work particularly well in our case mostly for the fact that i) there
are projection effects as we consider 2-dimensional clustering, ii) we
probe here more non-linear scales ($k_{max} \sim 3000$). In fact this
form of the covariance matrix is known to work until
$k^{3D}<0.8 $ hMpc$^{-1}$ while in our case $k^{3D}_{max} \sim 7$ hMpc$^{-1}$
for $\bar{z}=0.7$. Thus we propose a generalization introducing a scale dependent
$\sigma_{min} = \sigma_{min}(k)$.
Then Ansatz for the covariance matrix becomes:
\beq \label{Cov}
\textrm{Cov}_{ij} =  \delta_{ij} \frac{2 \lp P(k_i)+\frac 1 {\bar
    n}\rp^2}{N_{k_i}} +
\sigma_{min}(k_i)\sigma_{min}(k_j)P(k_i)P(k_j).
\enq
%In order to have an estimation of $\sigma_{min}^{2}(k)$,
% we write the
%leading term of the non-Gaussian part of the covariance matrix
%of the lognormal field as \citep[see e.g.][]{Takahashietal2014} :
%\beq
%\begin{split}
%&T_{ij} = 2P(k_{i})P(k_{j})(P(k_{i})+P(k_{j})) \\
%& + (P(k_{i})+P(k_{j}))^{2}
%[P(|k_{i}+k_{j}|)+P(|k_{i}-k_{j}|)].
%\end{split}
%\label{eq:cov2D}
%\enq
%To compare to our measurements, we need to shell averaged this expression (which affects only
%the last term in square brackets). 
%For simplicity let consider the
%diagonal terms for which $|k_{i}| \simeq |k_{j}| \simeq k$, then, $|k_{i} \pm k_{j}|^{2}
%= 2k^{2}(1 \pm \cos(\theta))$.
%Assuming an angular power spectrum of the form: 
%\begin{equation*} 
%P(k)=
%\begin{cases} 
%P(0) & \text{if $k=0$} \\ 
%A_{s}k^{n} &\text{if $k \neq 0$}
%\end{cases} 
%\end{equation*}

% ------------------- JC  8 March 15

In order to estimate of $\sigma_{min}(k)$, we proceed as follows. The leading term of the trispectrum of the lognormal field is given by
\citep[see e.g.][]{Takahashietal2014} :
\beq
\begin{split}
&T(\veck_i,-\veck_i,\veck_j,-\veck_j) = 2P(\veck_{i})P(\veck_{j})(P(\veck_{i})+P(\veck_{j})) \\
& + (P(\veck_{i})+P(\veck_{j}))^{2} [P(|\veck_{i}+\veck_{j}|)+P(|\veck_{i}-\veck_{j}|)].
\end{split}
\label{eq:cov2D}
\enq
To obtain the spectrum covariance matrix, we need to average this expression, summing over all Fourier modes $\veck_i$ and $\veck_j$ in the corresponding bins of shells of the spectrum estimator. Assuming the bin width is small enough, this averaging affects only the term in square brackets. In the limit of a large number of modes and infinitesimal bin width, it is the average with respect to the angle $\theta$ between $\hat \veck_i$ and $\hat \veck_j$. On the diagonal ($k_i = k_j = k$) it takes the form
\beq
\frac{1}{2\pi} \int_{-\pi}^{\pi} P \lp k\sqrt{2 + 2 \cos \theta} \rp d\theta.
\enq
It appears that for the relevant case of pure power law spectra $P \propto A_sk^{n}$ with realistic exponent, this integral diverges. We can correct this by separating the (in the real world finite and negligible) contribution of the background mode fluctuation $P(0)$ from the rest, which we still treat in the continuous limit. I.e. we write
\beq
 \frac{1}{2\pi} \int_{-\pi
  +\epsilon}^{\pi - \epsilon} A_{s} (2k^{2}(1+\cos(\theta)))^{n/2} d\theta.
\enq
The cutoff parameter $\epsilon = \arctan(1/n_{i})$, where $n_i =  k_i L/(2\pi) $ is such that the
integral starts at the first non-zero mode of the discrete Fourier modes associated to the grid.
%To avoid the divergence of this integral, we have used the
%discreteness of $k$ and extract the contribution from the zero mode.
The integration gives:
\beq
 \frac{2^{n+1}}{2\pi} P(k)
B_{\cos^{2}(\epsilon/2)} (1/2, (n+1)/2)
\enq
where B is the incomplete beta-function.
According to Equation~\ref{eq:cov2D}, we have on the diagonal of the
covariance of the galaxy field matrix:
\beq
\begin{split}
& \sigma_{min}^{2}(k_{i}) = T{ii}/P(k_{i})^{2} - \frac{2}{N_{k_{i}}} \\
& = 4 P(k_{i})\Big[ 1+
\frac{2^{n+2}B_{\cos^{2}(\epsilon/2)} (1/2, (n+1)/2)}{2\pi} \Big]
\end{split}
\label{eq:diag}
\enq
To estimate the slope $n$, we adjust a power-law to the 2-dimensional power
spectrum and use the best fit values presented in
Table~\ref{tab:values}.
\begin{table}
\centering
\caption{Best-fitting slope $n$ values and $\chi^{2}$/d.o.f derived
  using the fitting
  formula $P(k) = A_{s}k^{n}$ on the 2-dimensional field predictions in
  the four redshift bins.}
\begin{tabular}{cccc}
\hline
\hline
Redshift bin   &  $n$ & $\chi^{2}$/d.o.f\\
\hline
$0.2<z<0.4$ &   -1.34 & 0.04\\
$0.4<z<0.6$ &    -1.38 & 0.07\\
$0.6<z<0.8$ &    -1.44 & 0.21 \\
$0.8<z<1.0$ &    -1.63 & 0.21 \\
\hline
\hline
\end{tabular}
\label{tab:values}
\end{table}
\begin{figure*}
 \begin{center}
  \begin{tabular}{c@{}c@{}}
      \includegraphics[width=5.5cm]{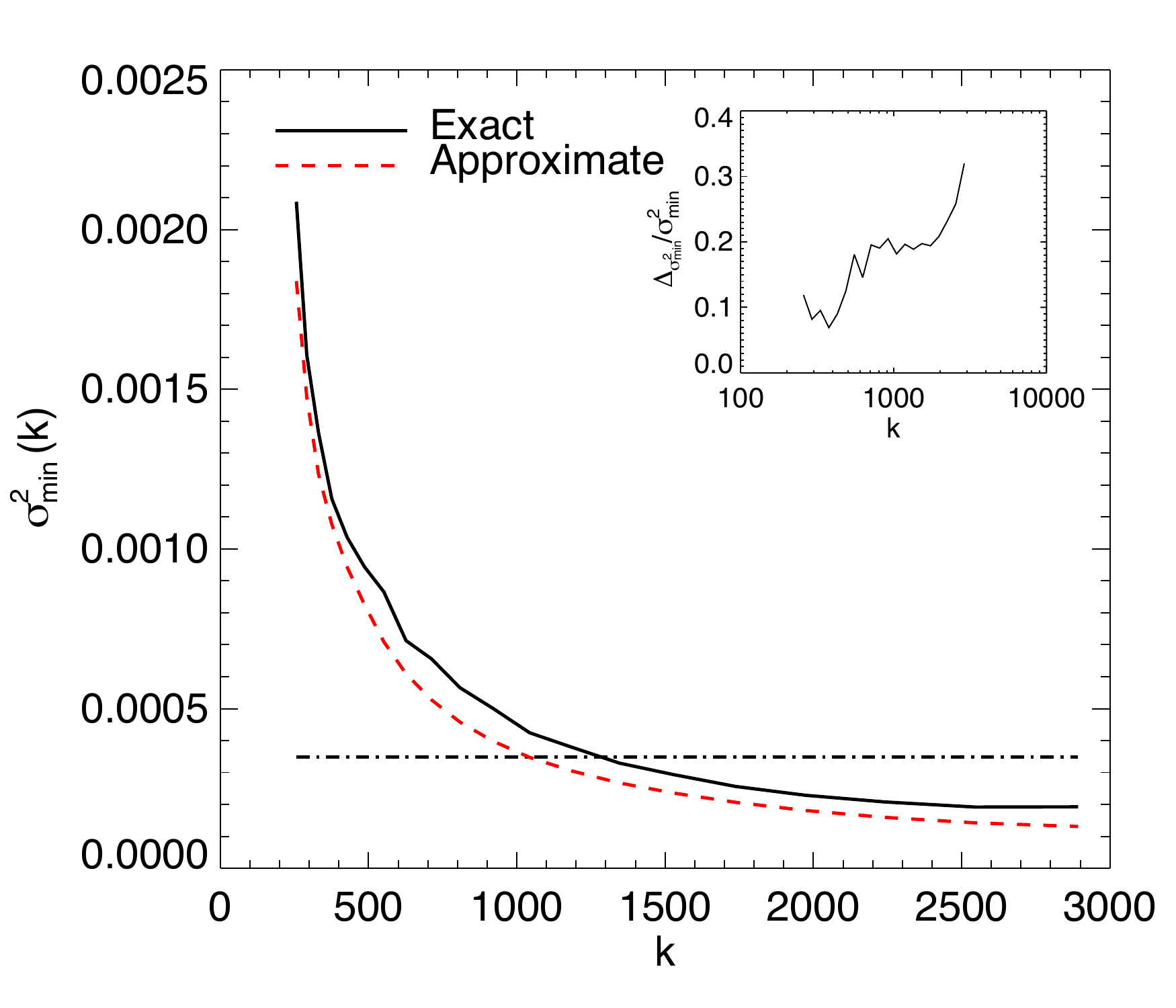}
      \includegraphics[width=5.8cm]{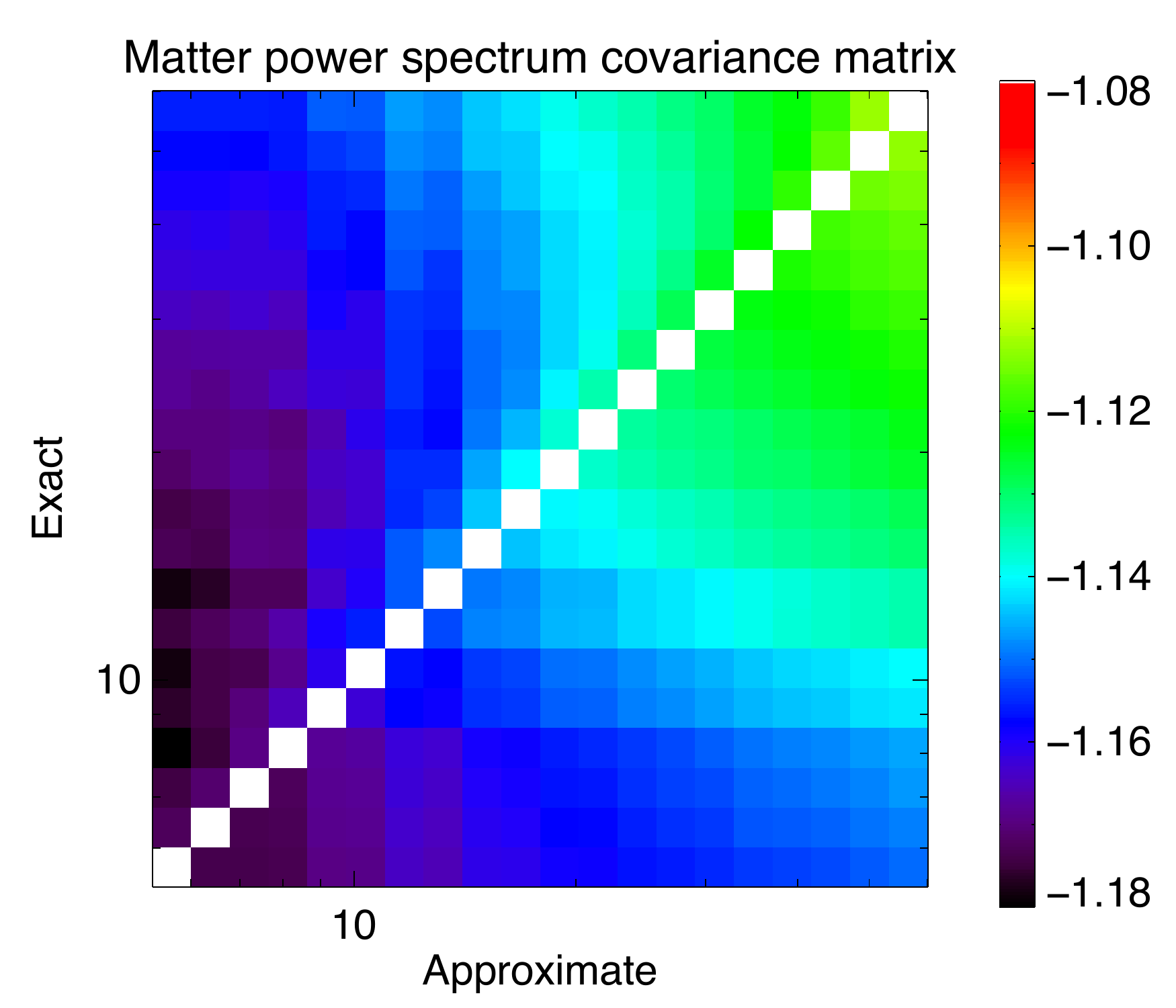}
      \includegraphics[width=5.5cm]{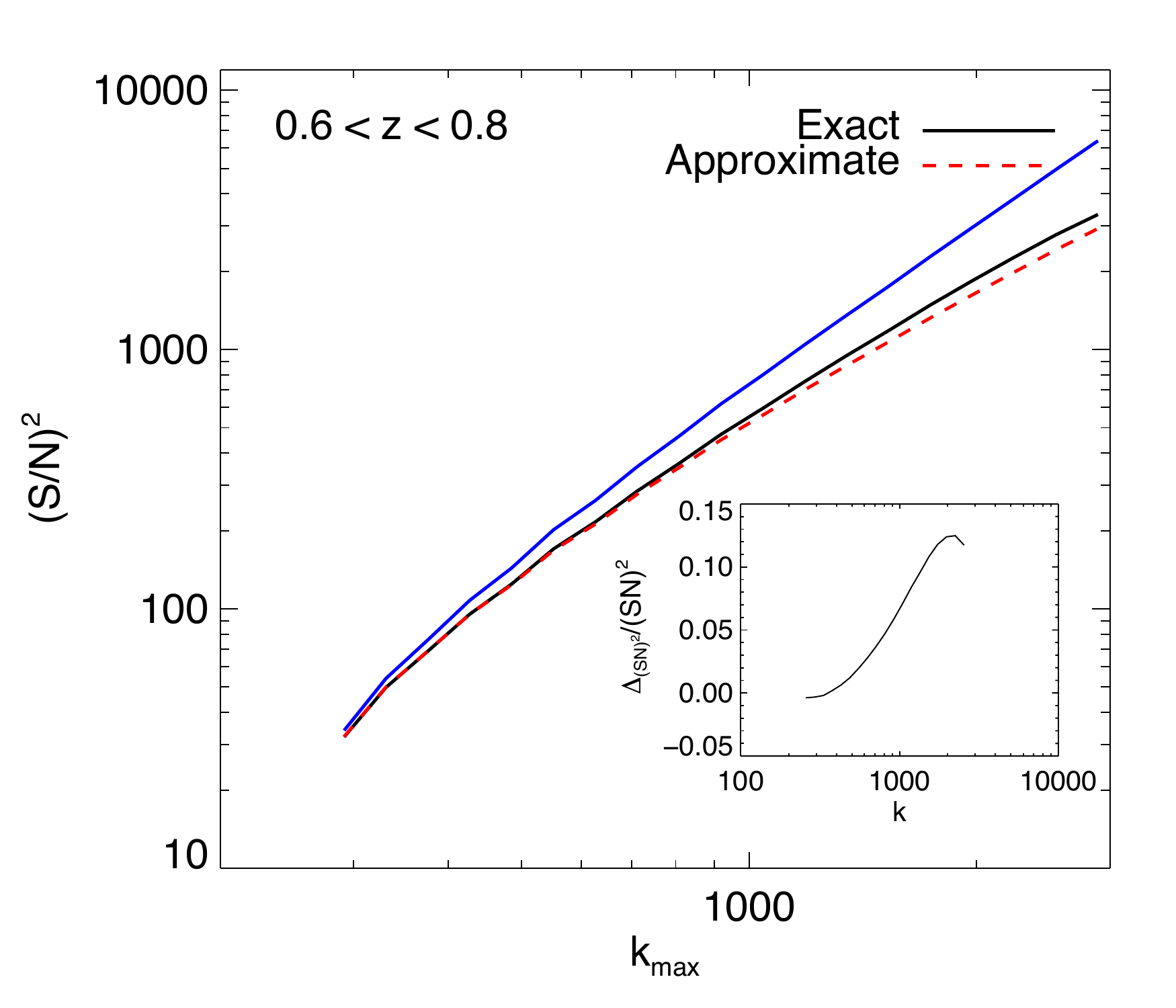}
      \end{tabular}
   \caption{ \label{figure:gcov} On the left panel, the comparison
between the predicted value of $\sigma_{min}^{2}(k)$ using the exact
matter covariance
matrix predicted by simulations compared to Equation~\ref{eq:diag}
(red dashed line). As a comparison the dot-dashed line shows
the value given by Equation~\ref{eq:sigcst} from
\protect\cite{Carronetal14d}. Our approximation
reproduces well the shape of the true value within $30\%$ at high-$k$
and $10\%$ at low-$k$. The middle panel shows the
comparison between the exact (upper left corner) and the approximate
(lower right corner) values of the matter covariance matrix. The right
panel shows the squared cumulative
signals-to-noise obtained using approximation of Equation~\ref{Cov}
for the galaxy power spectrum covariance matrix compared to the exact
value. At our resolution, the accuracy is within $\sim 10\%$.}
\end{center}
   \end{figure*} 

The left panel of Figure~\ref{figure:gcov} shows the comparison
between the predicted value of $\sigma_{min}^{2}(k)$ using the exact
matter covariance
matrix. The red dashed line shows $\sigma_{min}^{2}(k)$ given by
Equation~\ref{eq:diag} and as a comparison the dot-dashed line shows
the value given by Equation~\ref{eq:sigcst}. Our approximation
reproduces well the shape of the true value within $30\%$ at high-$k$
and $10\%$ at low-$k$.
The middle panel of Figure~\ref{figure:gcov} illustrates the
comparison between the exact and the approximate covariance matrix
while the right panel shows that the squared cumulative
signals-to-noise agree within $\sim 10\%$.

\section{Information content}
\label{sec:infocontent}

We now have all the ingredients to quantify the Fisher
information content of the $A^{\ast}$-power spectrum for cosmological
parameters (which in the shot-noise free regime is very close to the
total information).
We can also compare the cosmological information content of the galaxy power
spectrum to that of $A^{\ast}$.
Our analytical model only requires on the prediction of the galaxy power
spectrum $P$ and of the shot-noise level of the considered survey
through $\bar{N}$. Thus, for a given observation, we can forecast analytically the constraints on cosmological parameters extracted from the clustering of the underlying random field.
\newline
\newline
Given a set of parameters ${\alpha, \beta, ...}$, the Fisher matrix of
the spectrum is defined as:
\beq
F_{\alpha \beta} = \sum_{k_i,k_j < k_{max}} \frac{\partial
  P(k_i)}{\partial \alpha} [\textrm{Cov}_{ij}]^{-1} \frac{\partial P(k_j)}{\partial \beta}
\enq
where the covariance matrix is given by Equation~\ref{Cov}.
The inverse of the Fisher matrix
corresponds to the covariance of the posterior distribution of the
parameters that could be obtained given the error bars one has on the data.
It means that the larger the value of a Fisher matrix coefficient is, the smaller the
variance becomes, and therefore, the tighter the constraint on the
parameter.

\begin{figure*}
  \begin{center}
  \begin{tabular}{c@{}c@{}}
      \includegraphics[width=0.5\textwidth]{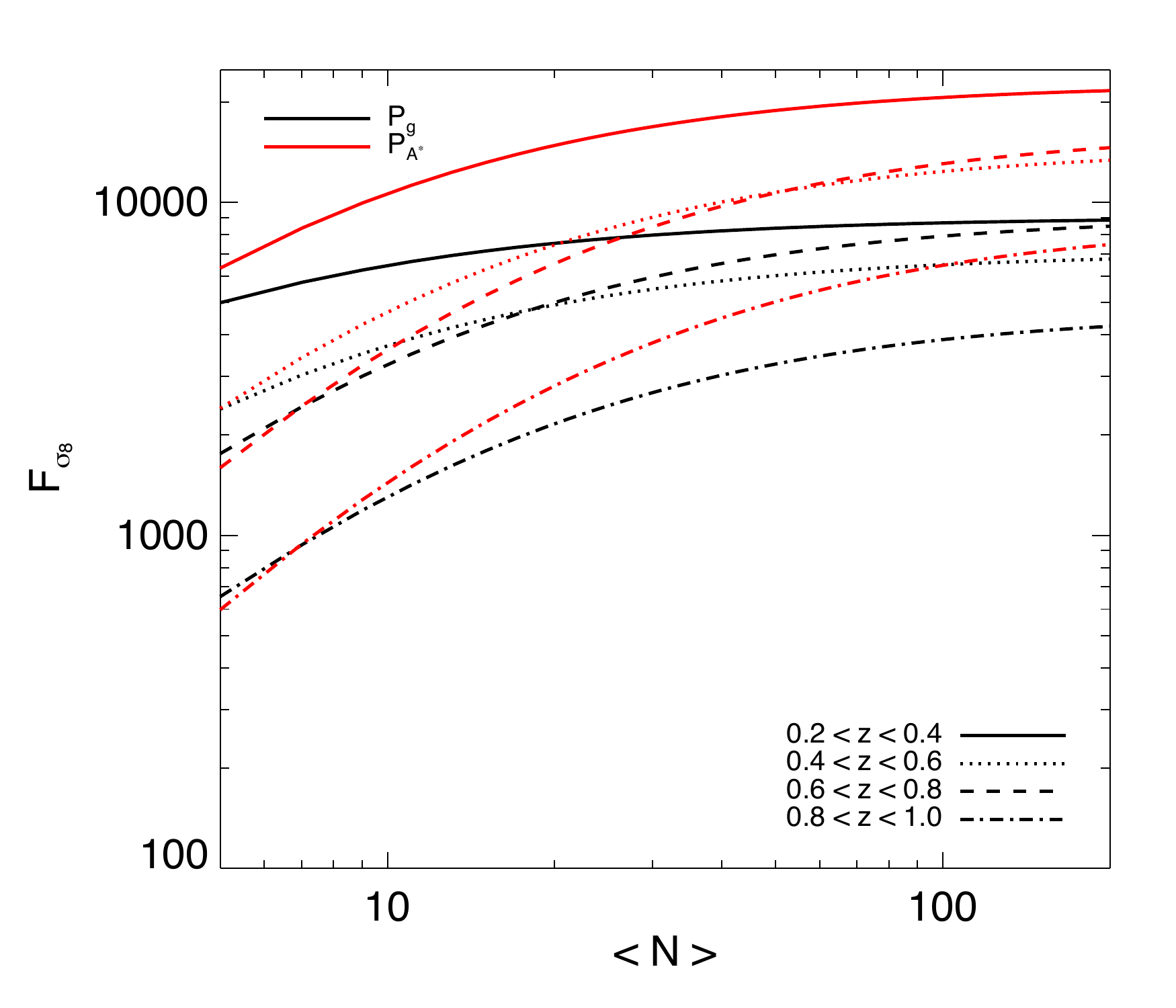}&
      \includegraphics[width=0.5\textwidth]{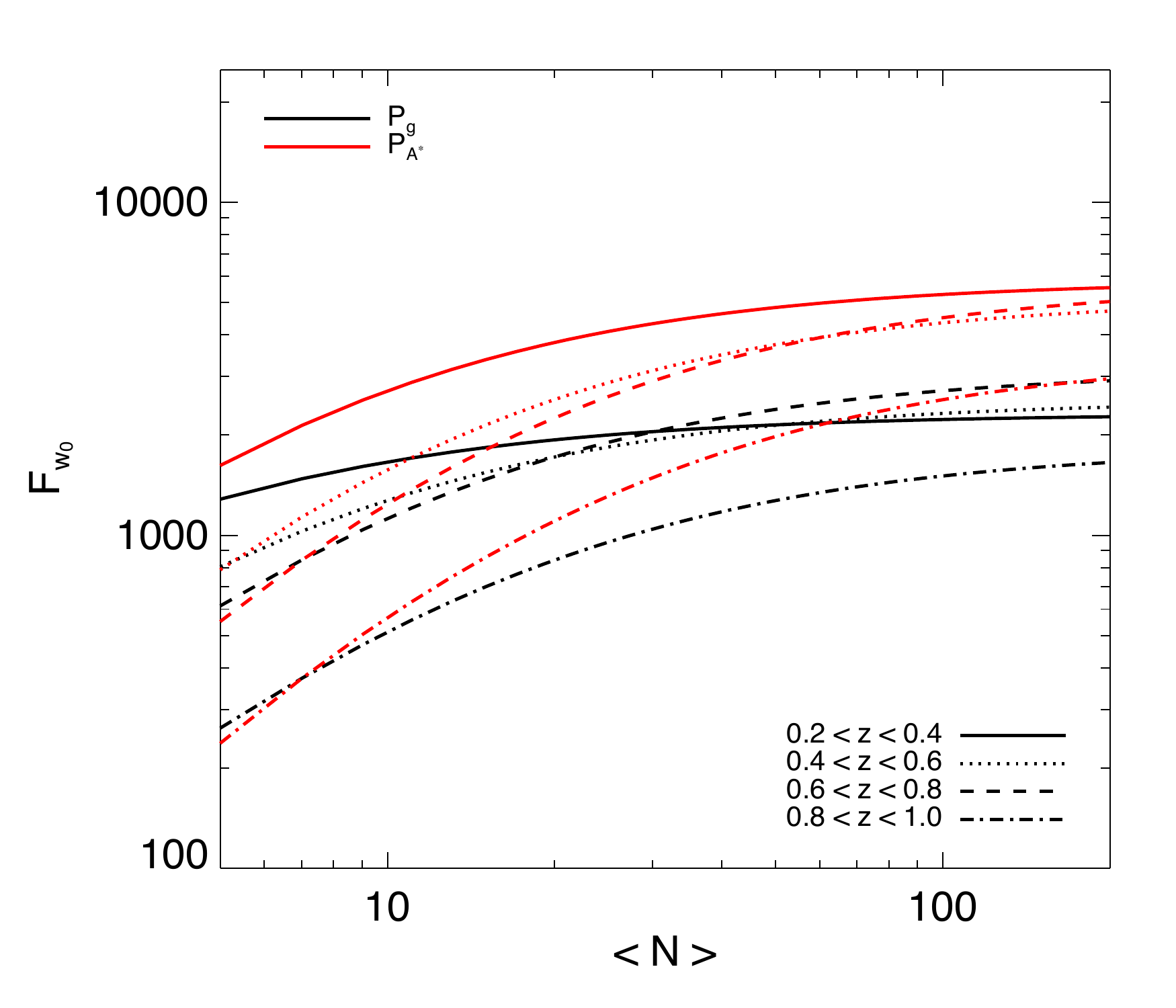}
\end{tabular}
    \caption{\label{figure:infopara} Fisher information of the
      $A^{\ast}$-(red lines) and galaxy power
spectrum (black lines) a function of the shot-noise level (through the
sampling rate
$\bar{N}$) for the two cosmological parameters $\sigma_{8}$ (left
panel) and
$w$ (right panel) in the four redshift bins $0.2<z<0.4$,
$0.4<z<0.6$, $0.6<z<0.8$ and $0.8<z<1.0$.
For $\bar{N} > 5-7$, we see that the non-linear transform $A^{\ast}$ performs
better than the galaxy power spectrum over the whole range of number
densities and redshifts to extract information on cosmological parameters.
It can also be seen that the $A^{\ast}$-power spectrum is more powerful at low redshifts where the
non-linearities are stronger, but also for dense survey (i.e for large
values of $\bar{N}$).}
\end{center}
    
\end{figure*} 

The information content from $A^{\ast}$ is given by:
\beq
F^{A^{\ast}}_{\alpha \beta} = \sum_{k_i,k_j < k_{max}} \frac{\partial
  P_{A^{\ast}}(k_i)}{\partial \alpha}
[\textrm{Cov}^{A^{\ast}}_{ij}]^{-1}\frac{\partial P_{A^{\ast}}(k_j)}{\partial
  \beta}.
\label{eq:Fas}
\enq
Thus Equation~\ref{eq:biasfinal} leads to
\beq
\textrm{Cov}^{A^{\ast}}_{ij} = \frac{2b^{4}}{N_{k}}
\Big(P(k_i)+\frac{1}{\bar{n}}\Big) \Big(P(k_j)+\frac{1}{\bar{n}}\Big) \delta_{ij}.
\enq
Equation~\ref{eq:Fas} then becomes:
\beq
F^{A^{\ast}}_{\alpha \beta} = \frac{1}{2} \sum_{k< k_{max}}
\frac{\partial \ln P_{A^{\ast}}}{\partial \alpha} N_{k} \frac{\partial
  \ln P_{A^{\ast}}}{\partial \beta}
\enq
with
\beq
\begin{split}
& \frac{\partial \ln P_{A^{\ast}}(k)}{\partial \alpha} =
\frac{\partial \ln b^{2}}{\partial \alpha }+ \frac{\partial \ln
P_{g}(k)}{\partial \alpha} \\
& = \frac{\partial \ln b_{A}^{2}}{\partial \alpha }+ \frac{\partial
P(k)}{\partial \alpha}\frac{1}{P(k)+1/\bar{n}}.
\end{split}
\enq
The bias coming from the $A^{\ast}$ mapping is fixed by the
fiducial values of HOD and cosmology and thus
does not carry a cosmological dependence.
\newline
\newline
Finally:
\beq
\begin{split}
& F^{A^{\ast}}_{\alpha \beta} = F^{G}_{\alpha \beta} + \frac{\partial \ln b_{A}^{2}}{\partial \alpha} \frac{\partial
  \ln b_{A}^{2}}{\partial \beta} (S/N)^{2}_{G} \\
& + \frac{\partial \ln
  b_{A}^{2}}{\partial \alpha}F_{\beta, \ln A_{z}}^{G} + \frac{\partial \ln
  b_{A}^{2}}{\partial \beta}F_{\alpha, \ln A_{z}}^{G}
\end{split}
\label{eq:infocontent}
\enq
where:
\beq
\begin{split}
 F^G_{\alpha \beta} =  \int d \ln k\: w(k)  \frac{\partial  \ln P(k)}{\partial \alpha}\frac{\partial  \ln P(k)}{\partial \beta}.
\end{split}
\enq
with
\beq \label{wk}
w(k) = \frac{V}{2}\frac{k^2}{2\pi} \Big(\frac{\bar{n}P(k)}{\bar{n}P(k) + 1}\Big)^{2}.
\enq
corresponding to the usual formula from \cite{Tegmark97}.
We have replaced the discrete sums with integrals using the fact that
the number of modes $N_{k}$ is approximately the surface of the shell used for the
bin averaging divided by the distance element between two discrete
modes. With our convention:
\beq
N_{k} \simeq V\frac{2\pi kdk}{(2\pi)^{2}}
\enq
Moreover, in Equation~\ref{eq:infocontent}, by analogy to
\cite{Carronetal14d}, we have introduce a nonlinear amplitude parameter
$\ln A_{z}$ defined such as $ \partial_{\ln A_{z}} P(k) = P(k)$. This
parameter corresponds to the initial amplitude $\sigma_{8}^{2}$ in the
linear regime and at $z=0$.
We further define the Gaussian signal to noise as:
\beq \label{SNG}
\lp S/N\rp^2_{G} = \int d \ln k \frac{V}{2}\frac{k^2}{2\pi}.
\enq
corresponding to a case without shot-noise.
The derivatives are estimated numerically using the \texttt{CosmoPMC} package.
 
The panels of Figure~\ref{figure:infopara} show the $A^{\ast}$- and galaxy power
spectrum Fisher information as a function of
$\bar{N}$  for the two cosmological parameters $\sigma_{8}$ (left
panel) and
$w$ (right panel) in the four redshift bins $0.2<z<0.4$,
$0.4<z<0.6$, $0.6<z<0.8$ and $0.8<z<1.0$. We consider values of
$\bar{N}>5$ where $A^{\ast}$ is expected to start to perform better.
We can clearly see that for small $\bar{N}$, the shot-noise erases
information present in the underlying random field. As previously seen in
\cite{Wolketal14}, the analytical approach developed in this work
reproduces well the general trends expected for the non linear
transform $A^{\ast}$: i) for $\bar{N} > 5-7$, $A^{\ast}$ performs
better than the galaxy power spectrum over the whole range of number
densities and redshifts and thus could be used to unveil the otherwise hidden
information, ii)  the use of the $A^{\ast}$-power spectrum to extract the information 
is more powerful at low redshifts where the non-linearities are stronger, and iii)
our observable is more efficient for dense survey (i.e for large
values of $\bar{N}$).

\begin{figure*}
  \begin{center}
  \begin{tabular}{c@{}c@{}}
      \includegraphics[width=0.5\textwidth]{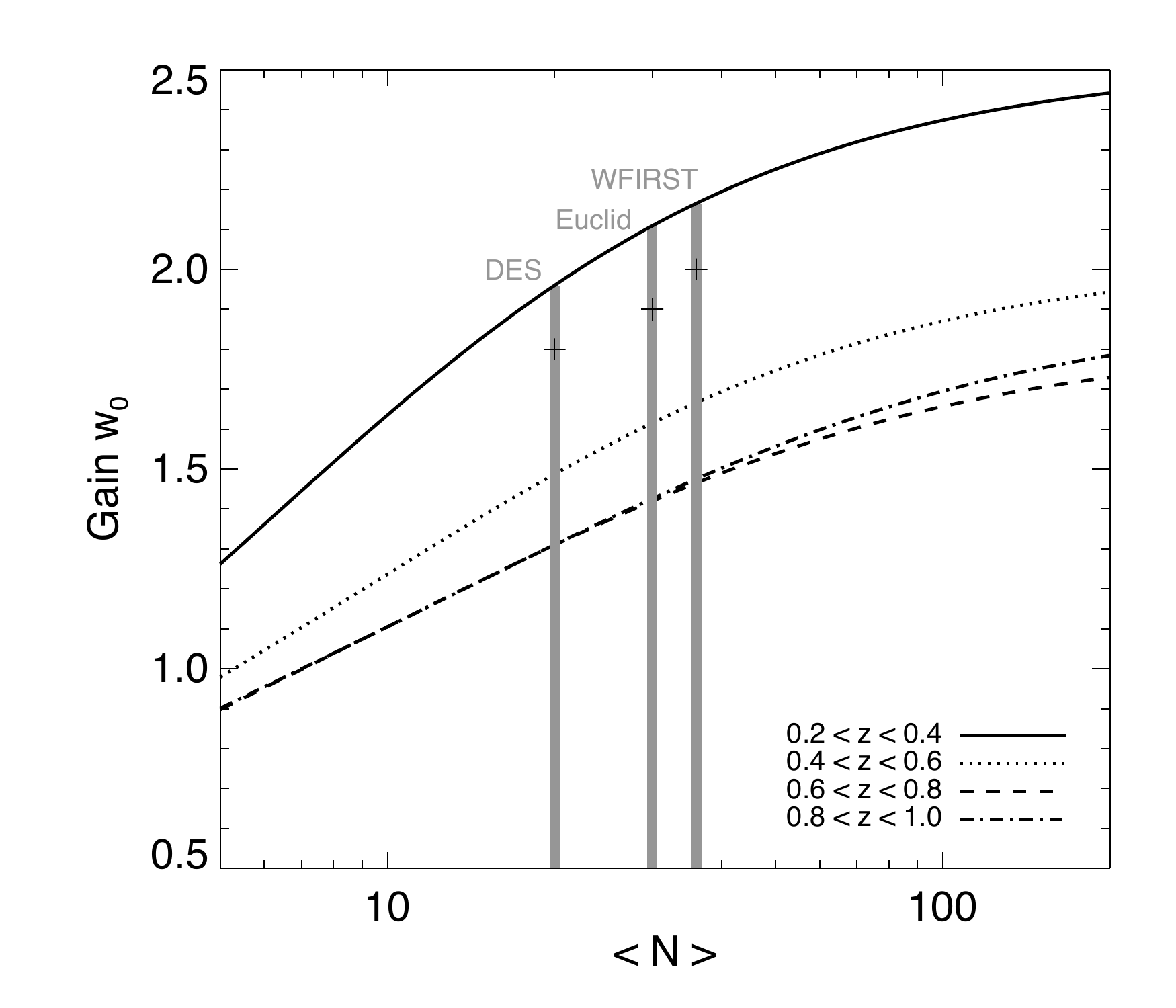}&
      \includegraphics[width=0.5\textwidth]{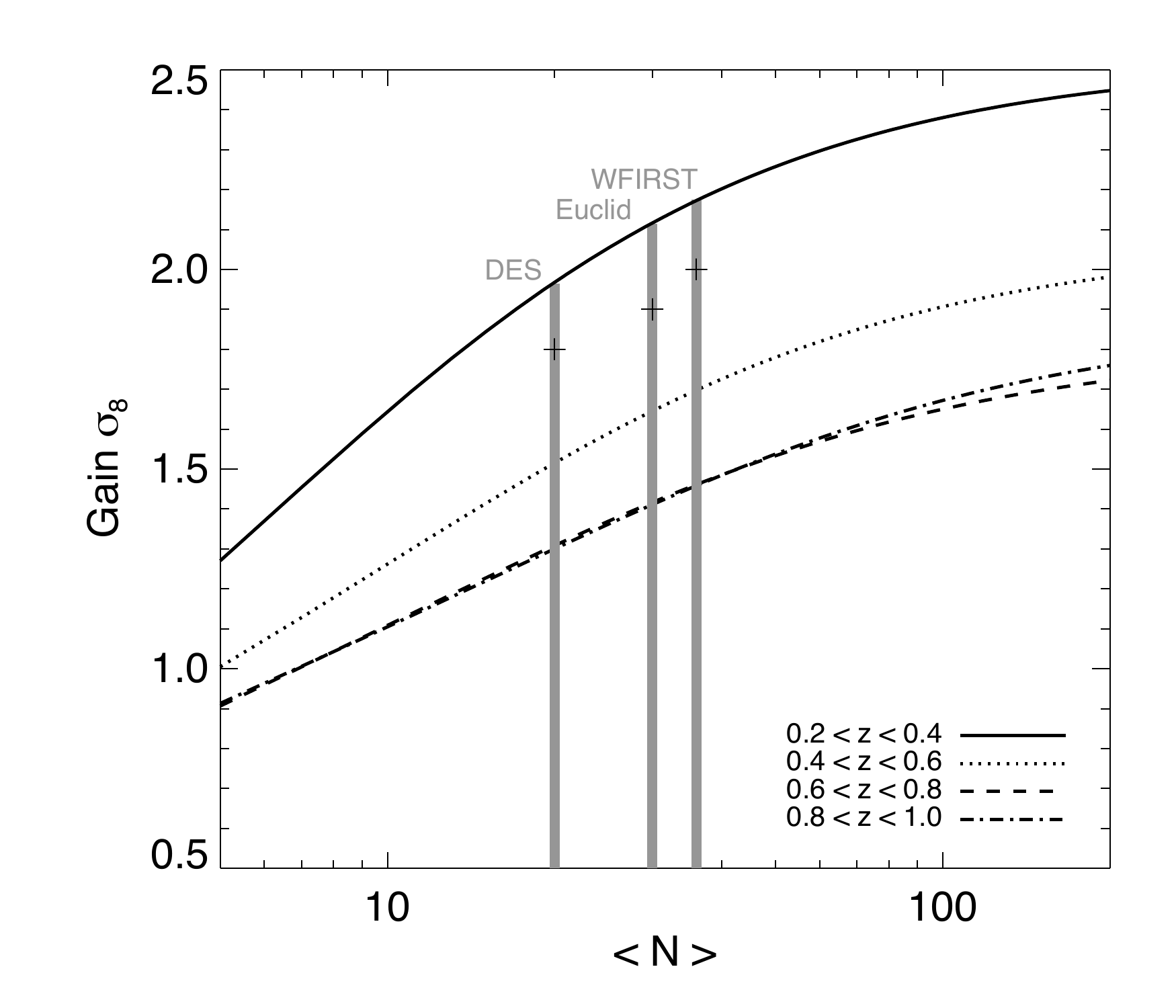}
\end{tabular}
    \caption{\label{figure:gain} Information gains on the
  cosmological parameters
  $w_{0}$ (left panel) and $\sigma_{8}$ (right panel) using the
  $A^{\ast}$ power spectrum
  instead of the galaxy power spectrum as a function of the shot-noise
  level in the survey (through the
sampling rate
$\bar{N}$). We recover the gain predicted numerically in
\protect\cite{Wolketal14} (black crosses) which was about a factor of 2, especially at low
redshifts and for dense surveys. Illustrated by the vertical lines,
the values of $\bar{N}$ for different upcoming surveys in the first
redshift bin $0.2<z<0.4$.}
\end{center}
    
\end{figure*} 

To compare in a quantitative way our results with the previous
forecasts of \cite{Wolketal14}, Figure~\ref{figure:gain} shows the predicted improvement in the information on $w_{0}$ (left panel) and $\sigma_{8}$ (right
panel) as a function of redshift and the
survey shot-noise level. The quantity plotted is the ratio
between the galaxy and $A^{\ast}$-power spectrum Fisher matrix elements. In that sense,
it represents the expected information gain using the
non-linear transform $A^{\ast}$ instead of the power spectrum. The
simplest interpretation of this gain is an effective gain in survey
area. We also show for illustrative purposes, the values of $\bar{N}$
for different upcoming surveys in the first redshift bin where the
gain is known to be the highest
\citep[see][for details]{Wolketal14}.

This analytical approach reproduces better than $20\%$ the
expected gain for the
two parameters $\sigma_{8}$ and $w_{0}$.  Qualitatively, the achievable gain is about a factor of 2, especially at low
redshifts and for dense surveys.  We conclude that the analytical
model developed here using the matter power spectrum at a redshift $z$
and the number density of the survey, is able to predict the constraints on cosmological parameters from galaxy clustering with reasonable precision.

\section{Discussion}
\label{sec:discuss}

It has been known that non-linear transforms help to capture more
efficiently the information encoded in the matter density field. The
notion of {\em sufficient statistics} \citep{Carronetal13} has emerged
as the optimal transformation that extracts all cosmological
information. In the case of a discrete galaxy field the new observable $A^{\ast}$ was constructed \citep{Carronetal14a}.
\cite{Wolketal14} have forecasted using a numerical approach the
expected improvement on constraints beyond that of using the galaxy
power spectrum on the latest CFHTLS data set as well as on upcoming large wide-field surveys; for the former, the forecast agreed well with the actual gain realized when calculating the $A^{\ast}$ power spectrum.
In this work, we have developed an analytical approach that captures
the statistics of $A^{\ast}$ to the point that we could accurately forecast
the best achievable constraints on cosmological
parameters as a function of the survey density. The forecast improvement is consistent with previous, more tedious, numerical calculations at the $20\%$ level at worst (or $10\%$ for error bars).
\newline
\newline
We have presented an Ansatz for the bias between the galaxy
and $A^{\ast}$-power spectra, and demonstrated its
accuracy compared to the previous numerical approach.
We showed that the dependence of the bias on cosmology is crucial for endowing
$A^{\ast}$ with the ability to recapture the hidden information from the field.
In addition, we proposed a diagonal form for the $A^{\ast}$ power spectrum
covariance matrix and showed that it is accurate at the $5\%$ level.
\newline
\newline
In order to compare with the standard method of extracting cosmological parameters
from the galaxy power spectrum, we have provided and explored the accuracy of an
analytical Anstatz for the projected power spectrum 
covariance matrix. Based on a
generalization of \cite{Carronetal14d}, we were able to reproduce
squared cumulative signals-to-noise of the matter field within $10\%$.
\newline
\newline
Although our analytical framework contains
a fair number of  approximations, we have demonstrated that our forecasts
are reliable at least within $20\%$ even at the most non-linear scales
we probed. Moreover, our method includes all the non-Gaussian
effects (super survey modes, trispectrum, discreteness) and thus it is
expected to be more accurate than the standard Gaussian forecasts entirely ignoring such effects.
In addition, the approach has also provided new insights and a
 deeper understanding of the cosmological information content of the galaxy clustering.
\newline
\newline
Finally, we predicted the best achievable constraints on the
cosmological parameters: $\sigma_{8}$ and $w_{0}$ as a function of the
shot-noise level in the survey.
We were able to recover, the predictions from
\cite{Wolketal14} using a large ensemble of numerical simulations, and found that the
gain on the information using the $A^{\ast}$-power spectrum translates into
factor of 2 gain approximately, especially at low redshifts and for dense surveys.
\newline
\newline
The promise of $A^{\ast}$ for improving cosmological  constraints from future surveys has been clear for a while. However, until now, the prediction of its power spectrum involved a large number of numerical simulations, a disadvantage when used in an MCMC sampling framework to fit cosmological parameters. Likewise, the corresponding covariance matrices also needed massive number of simulations. The present work provides a convenient and accurate short cut, that can be used at least for forecasting purposes, and it has the potential of speeding up MCMC sampling as well.
The present approximations have been tested for 2-dimensional projected surveys, but similar developments can be carried out for  3-dimensional
surveys as well. Previous attempts have been made using dark matter
simulations, however, it is worth mentioning that the information gain is
volume dependent and changes with respect to a local or global
description \citep[i.e if we consider density fluctuations defined with respect to the
local observed density or not, see][]{Carronetal14d}. 
\cite{Neyrincketal09} using the local density field from
the 500 h$^{-1}$Mpc Millenium simulation \citep{Springel05} found a
factor of $\sim10$ improvement on the information on the $(S/N)^{2}$
(corresponding approximately to $\ln(\sigma_{8}^{2})$ unmarginalized
over the other cosmological parameters) using
sufficient statistics. \cite{Neyrinck11} doing the same analysis on
the Coyote Universe \citep{Heitmannetal09, Heitmannetal10} which have
a box size of 1300 Mpc, found then an improvement of $\sim 15$. More
recently, \cite{Wolketal15b} considering the contraints on neutrino
mass from the DEMNUNI simulation of volume $V = 8$ h$^{-3}$Gpc$^{3}$
using both the power spectrum and sufficient statistics found a factor
$\sim 8$ improvement on the information. A similar framework to the present for 3-dimensional 
surveys including the effects of redshift space distortions both on
the power spectra and on covariance matrices would be desirable for
applications, and are left for future work.

%\clearpage%
\newpage
The authors acknowledge NASA grants NNX12AF83G and NNX10AD53G for support.
\newline
\indent
Part of this work was based on observations obtained with MegaPrime/MegaCam, a joint project of CFHT and CEA/IRFU, at the Canada-France-Hawaii Telescope (CFHT) which is operated by the
National Research Council (NRC) of Canada, the Institut National des
Science de l'Univers of the Centre National de la Recherche
Scientifique (CNRS) of France, and the University of Hawaii. This work
is based in part on data products produced at Terapix available at the
Canadian Astronomy Data Centre as part of the Canada-France-Hawaii
Telescope Legacy Survey, a collaborative project of NRC and CNRS.

%------------------------------
%\appendix

\bibliographystyle{mn2e}
\bibliography{Notes.bib}

\end{document}